\documentclass[oneside]{article}
\usepackage{natbib}
\usepackage{graphicx}
\usepackage[sc]{mathpazo}
\usepackage{float} 
\usepackage[pdfproducer={Dikun},pdfcreator={Dikun},pdfauthor={Dikun}]{hyperref}
\usepackage{booktabs} 
\usepackage{fancyhdr} 
\usepackage{paralist} 
\usepackage[hang, small,labelfont=bf,up,textfont=it,up]{caption}
\usepackage[hmarginratio=1:1,top=32mm,columnsep=20pt]{geometry} 
\usepackage{multicol}

\usepackage{abstract}

\usepackage{titlesec} 
\renewcommand\thesection{\Roman{section}} 
\renewcommand\thesubsection{\Roman{subsection}} 
\titleformat{\section}[block]{\large\bfseries\scshape\centering}{\thesection.}{1em}{} 
\titleformat{\subsection}[block]{\large\bfseries}{\thesubsection.}{1em}{} 

\title{\vspace{20mm}\fontsize{22pt}{10pt}\selectfont\textbf{Parameterization of geophysical inversion model using particle clustering}
\thanks{Originally a course project report in April 2009 titled: Preliminary study of elementary inversion and d-norm.}} 

\author{
\large
\textsc{Dikun Yang}\\[2mm] 
\normalsize University of British Columbia \\ 
\normalsize \href{mailto:yangdikun@gmail.com}{yangdikun@gmail.com} 
\vspace{-5mm}
}
\date{}

\begin{document}

\maketitle


\begin{abstract}

\noindent 
This paper presents a new method of constructing physical models in a geophysical inverse problem, when there are only a few possible physical property values in the model and they are reasonably known but the geometry of the target is sought. The model consists of a fixed background and many small ``particles'' as building blocks that float around in the background to resemble the target by clustering. This approach contrasts the conventional geometric inversions requiring the target to be regularly shaped bodies, since here the geometry of the target can be arbitrary and does not need to be known beforehand. Because of the lack of resolution in the data, the particles may not necessarily cluster when recovering compact targets. A model norm, called distribution norm, is introduced to quantify the spread of particles and incorporated into the objective function to encourage further clustering of the particles. As proof of concept, 1D magnetotelluric inversion is used as example. My experiments reveal that the particles, starting from a fully scattered distribution, are able to move towards the actual target location; the quality of recovery depends on whether there is enough material (vertical conductance in 1D) in the particles to build the target; and the use of distribution norm can help produce tightened clustering. When the inversion struggles to fit the data, it may indicate that the prior information about the particles' conductivity and size are incorrect.

\end{abstract}


\section{Introduction}
Geophysical inversions recover the unknown distribution of physical property of the earth using the observed data. The distribution, usually referred to as ``model'', can be parameterized differently depending on the knowledge of the earth we may have prior to the inversion. In the most generic model, the earth is discretized into many ``pixels'', each of which has a physical property value that is searched by the inversion; the model parameters are the values of the pixels \cite[]{constable1987occam}. Another well-established parameterization is to assume the recovered target to be some regularly shaped bodies, like sphere, plate, prism, etc.; the model parameters therefore include the physical property of the target and/or the geometric parameters of the body \cite[]{keating1990,smith2002moments}. The pixel inversion is highly flexible and works with arbitrarily complicated earth models, but its inversion often has high degree of freedom as every pixel is an unknown. This high degree of freedom can bring severe non-uniqueness and often a large amount of computation in inversions. The geometric inversion, having reduced number of model parameters, is usually much faster in computation thanks to the analytic or semi-analytic solutions, and provides robust model recovery. However, it requires the target can be represented by some simple bodies and the geometry of the body must be reasonably known before the inversion. This paper presents a new method of constructing the model using both the concept of pixel and geometric parameters: the earth is described by many freely moving elementary particles in a given background; the particles are assigned pre-determined physical properties; the model parameters are the locations of the particles.

The new approach is designed to work when specific information about the physical properties of the targets and background are available and desired to be incorporated into the inversion. Previously, this type of information can be dealt with under the framework of the pixel inversion by using upper and lower bounds \cite[]{musil2003discrete}, but one needs to know where the bounds should be applied over the modeling domain. The advantage of the particle clustering is that the particles are driven by the data to the locations where the physical property is required to be different from the background. The reduced model space, now spanned by the particles locations, makes the parameterization simple and the inversion robust.

If, due to the intrinsic non-uniqueness, the particle locations cannot be uniquely determined by the data, additional constraints are necessary to ensure the inversion results are geologically plausible. Since the conventional model norms that encourage ``small'', ``simple'' or smooth models in a pixel inversion do not apply to the location parameters. I therefore propose some model norms that measure the distribution of the particles and penalize heavily if the particles are widely spread. The model norms that constrain the particles' locations are collectively called ``distribution norm'' or ``d-norm''. I will show the use of d-norm calculated with two formulas using the variance and range.

As the first pass, I investigate the application of the particle clustering inversion in 1D magnetotelluric (MT) inversion problem because the forward solution can be rapidly computed. Many different scenarios are experimented to show the performance of the new parameterization if some prior information is incorrect and the earth contains multiple targets.

\section{Methods}

\begin{figure}[h]
  \centering
    \includegraphics[width=0.7\textwidth]{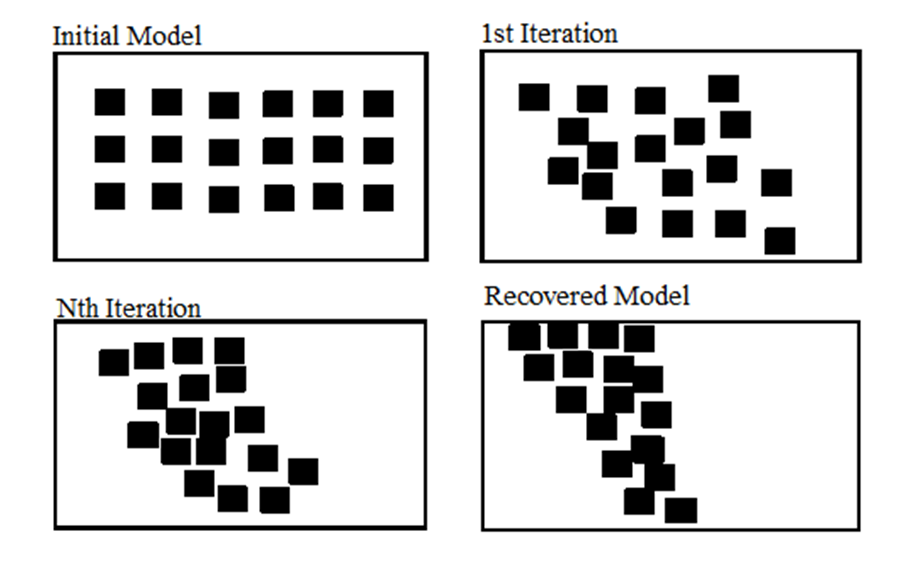}
  \caption{Particle clustering inversion in 2D.}
  \label{fig:clustering}
\end{figure}

The basic idea of the particle clustering inversion is illustrated in Figure~\ref{fig:clustering}, in which the locations of the particles are shown on the snapshots at four inversion iterations. Initially the background is assigned a known structure, for example, a uniform half-space for the simplest case. The particles are scattered over the entire domain. At each iteration, the inversion gradually moves the particles around in attempt to fit the observed data. Eventually the particles cluster to resemble the desired target and the model should reproduce the data if the physical property values are correctly assigned and there are enough particles in total volume. 

\section{1D Magnetotelluric Inversion Problem}

To verify the basic idea of particle clustering parameterization, I choose the 1D magnetotelluric (MT) inversion as example because the forward solution can be analytically computed and the frequency sounding has inherent depth resolution \cite[]{cagniard1953basic}. My experiments use 16 MT frequencies: 320, 160, 80, 40, 20, 10, 6, 3, 1.5, 0.75, 0.375, 0.1875, 0.09375, 0.04688, 0.02344, 0.01172 Hz. The data, for the convenience of visualization, are the apparent resistivity in $\Omega$m and phase in degree calculated from the impedance on the surface. 

The objective function minimized in the inversion is similar to the one used in most of regularized inversions
\begin{equation}
\label{eq:obj}
	\Phi = \Phi_d + \gamma \Phi_m,
\end{equation}
where $\Phi_d$ is the data misfit and $\Phi_m$ is the model norm, and $\gamma$ balances the relative importance of $\Phi_d$ and $\Phi_m$. The misfit $\Phi_d$ contains one term for the apparent resistivity and another for the phase. Because of the different dynamic ranges in the resistivity and phase, the apparent resistivity is re-parameterized by the common logarithm (log10) and then normalized by the observed apparent resistivity data. Both terms are based on L-2 norm. The data misfit is then
\begin{equation}
\label{eq:phid}
	\Phi_d = \Phi_{\rho} + \Phi_{\phi} = \left\Vert  \frac{ log(\rho^{obs}) - log(\rho^{pre}) }{ log(\rho^{obs}) } \right\Vert ^2 + \left\Vert  \phi^{obs}-\phi^{pre}  \right\Vert ^2,
\end{equation}
where the superscript $obs$ and $pre$ denote the observed and predicted data respectively. The formulation of the model norm will be discussed later.

The minimization of equation~\ref{eq:obj} requires the evaluation of its gradient, which involves the derivative of $\Phi_d$ and $\Phi_m$ with respect to the model parameters (particle locations). Unfortunately, $\Phi_m$ that measures the distribution of the particles may not necessarily be differentiable. As a result, I use the brute-force (perturbation) method to compute the gradient. Because of the new way of model parameterization, the model space is reduced to a few locations of the particles and the brute-force method is not too time-consuming.

\section{Non-regularized Inversions}

The first suite of experiments sets $\gamma$ in equation~\ref{eq:obj} to zero, so essentially the inversion clusters the particles using only the information from the data without any additional constraint. Two units are placed in the true model: a 1000 $\Omega$m uniform background and a conductive layer of 10 $\Omega$m. For 1D model, a building block particle is in fact a thin layer with fixed thickness. In my experiments, the size of a particle (thickness of a layer) is 100 m. The particles layers are allowed to overlap when moving, which makes the inversion more robust if too many particles are deployed at the beginning. 

\subsection*{Experiment 1: Perfect a priori information}
\begin{figure}[h]
  \centering
    \includegraphics[width=0.7\textwidth]{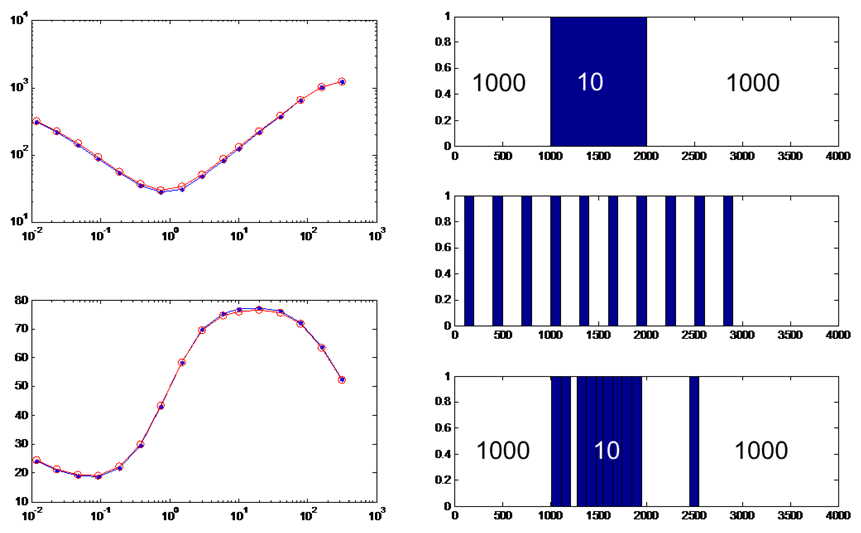}
  \caption{Experiment 1: perfect a priori information. Top-left: observed (blue) and predicted (red) apparent resistivity data at all frequencies. Bottom-left: observed (blue) and predicted (red) phase data at all frequencies. Top-right: the true resistivity model as a function of depth. Middle-right: the initial distribution of the particles (layers). Bottom-right: the model recovered by the particle clustering. The figures in other experiments follow the same layout.}
  \label{fig:simple}
\end{figure}

In this experiment, the target layer is at depth between 1000 and 2000 m. A synthetic data set is generated from the true model for the use in the inversion. I assume the exact resistivities of the background and the target are well known and the total number of layers is just enough to represent the true model. The experiment is shown in Figure~\ref{fig:simple}. Initially, the layers are placed with equal spacing from the surface to 3000 m deep to pretend no specific information about the target location is known. In this perfect scenario, the inversion is able to reasonably cluster the scattered layers to the correct location of the true target and reproduce the observed data very well. However, there is a small gap in the recovered conductive layer and a single layer left isolated at depth of 2500 m. Although the first-order structure is successfully reconstructed, those irregularities could make the inversion model less geologically plausible.

\subsection*{Experiment 2: Not enough particles}

\begin{figure}
  \centering
    \includegraphics[width=0.7\textwidth]{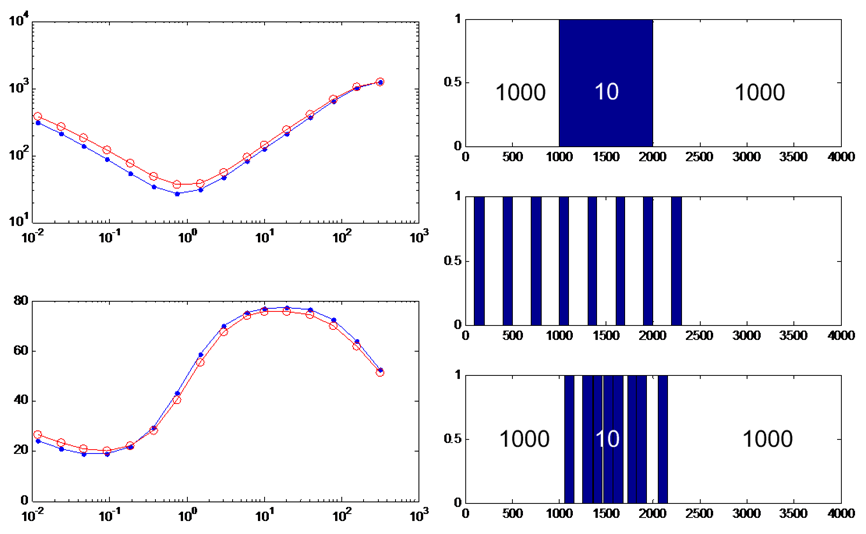}
  \caption{Experiment 2: not enough particles. }
  \label{fig:lesslayers}
\end{figure}

In practice, it is difficult to always have perfect information about the resistivity and total volume to initiate the particle clustering. In this experiment, I still use the same true model in Experiment 1, but assume not enough particles are deployed; the target resistivity is still correct. The true model needs 10 layers of 100 m thick, but only 8 layers are placed into the background. The experiment is shown in Figure~\ref{fig:lesslayers}. The recovered model shows that the particle layers are clustered at almost the correct location, although the small gaps between the layers are evident. I note the data fit, the best I can possibly get, is not as good as in Experiment 1. This may be an indication that the a priori information used to start the inversion could be incorrect. 

\subsection*{Experiment 3: Too many particles}

\begin{figure}
  \centering
    \includegraphics[width=0.7\textwidth]{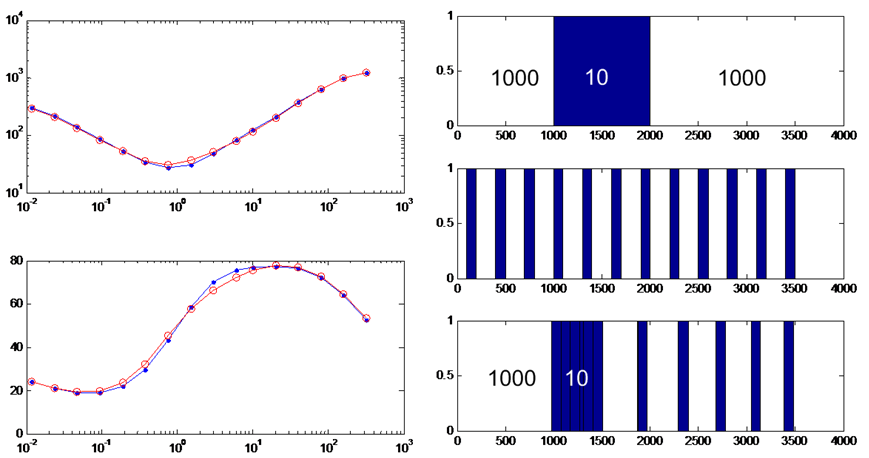}
  \caption{Experiment 3: too many particles. }
  \label{fig:morelayers}
\end{figure}

Another possible scenario of making incorrect input is having too many particles while the resistivities are still correct. This experiment is similar to Experiment 2, except 12 particle layers are placed into the background. The experiment is shown in Figure~\ref{fig:morelayers}. The recovered model predicts the observed data much better than in Experiment 2, showing that there are enough particles in terms of the total volume to reconstruct the true model. The top half of the target layer is very well recovered; however, the bottom of the target layer is rendered as some scattered layers. This indicates that the data set itself may not have the resolution for the target layer's bottom, which can also be inferred from Experiment 1. I also note there are 10.5 layers visible in the recovered model; the particles can effectively disappear by overlapping if some of the building materials are deemed unnecessary.

\subsection*{Experiment 4: Over-estimated resistivity}

\begin{figure}
  \centering
    \includegraphics[width=0.7\textwidth]{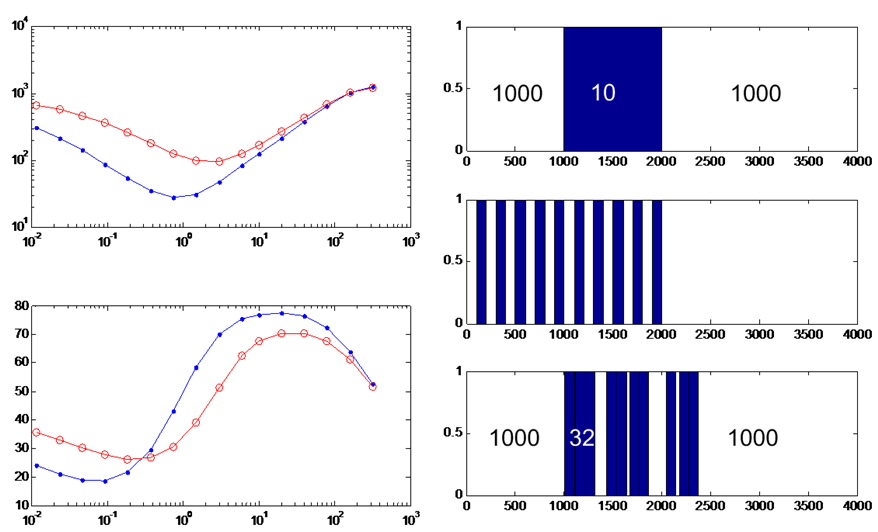}
  \caption{Experiment 4: over-estimated resistivity. }
  \label{fig:moreresistive}
\end{figure}

Now I test the scenario when the resistivity of the target is incorrectly guessed. In this experiment, I assume the resistivity of the target layer is over-estimated to 32 $\Omega$m from the true value 10 $\Omega$m. The number of particle layers is 10, which is just geometrically enough to build the target. The experiment is shown in Figure~\ref{fig:moreresistive}. Because of the lack of enough building materials (vertical conductance), the inversion ends up with a very poor data fit. Similar to Experiment 2, the bad misfit can indicate incorrect a priori information. The recovered model, certainly not perfect, does however provide some rough estimate of the target layer's location. 

\subsection*{Experiment 5: Under-estimated resistivity}

\begin{figure}
  \centering
    \includegraphics[width=0.7\textwidth]{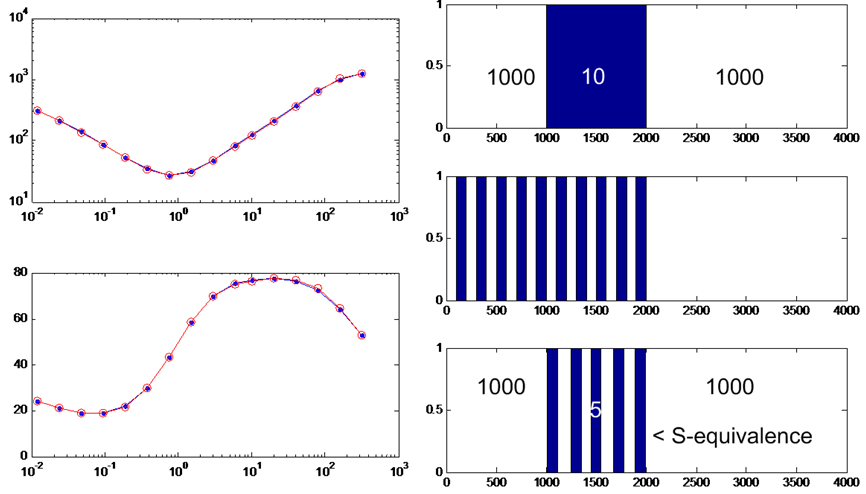}
  \caption{Experiment 5: under-estimated resistivity. }
  \label{fig:lessresistive}
\end{figure}

This experiment tests the scenario when the resistivity of the target layer is under-estimated to 5 $\Omega$m but the number of particles is correct. The experiment is shown in Figure~\ref{fig:lessresistive}. Under-estimated resistivity means there are excessive building materials. The inversion model shows that 5 of the 10 particle layers disappeared by overlapping; the remaining layers all together, distributed exactly within the depth range of the true target, have the same conductance as the original target layer. This experiment again shows the particle clustering is robust if the number of particles and/or the conductivity of the target are over-estimated at the beginning. 

\subsection*{Experiment 6: Multiple targets}

\begin{figure}
  \centering
    \includegraphics[width=0.7\textwidth]{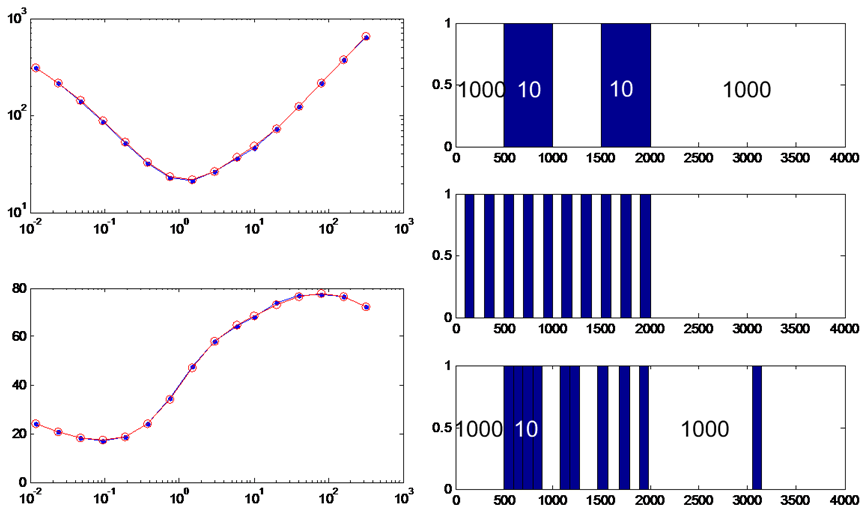}
  \caption{Experiment 6: multiple targets. }
  \label{fig:multilayers}
\end{figure}

In reality, knowing the actual number of targets underground can be sometimes difficult. This experiment uses two conductive layers (depth from 500 to 1000 m and from 1500 to 2000 m) in the true model, while still assuming a uniform distribution of the particles at the beginning. The experiment is shown in Figure~\ref{fig:multilayers}. The observed data have been fit very well. The top of the shallow target is well recovered by the clustering; however, its bottom and the boundaries of the deep target are not resolved, although the recovered model shows some hints for the lower boundary of the deep target.

\subsection*{Summary}
The preliminary results obtained from Experiment 1 - 6 has shown us:
\begin{compactitem}
 	\item Clustering some particles that have been assigned pre-determined properties is a viable way of constructing an inversion model. 
 	\item Perfect a priori information about the resistivity and the total volume of the target are crucial. If excessive conductive materials are deployed into the background, the inversion is able to ``hide'' the materials by overlapping the particles. If not enough conductive materials are available, the inversion may struggle to fit the observed data. 
 	\item The particle clustering, even with reduced model space, still has the non-uniqueness problem. In the 1D MT problem, the data do not have very good resolution to the locations of the particles at depth. Therefore, regularization is required to constrain the particle locations for more plausible models.
 \end{compactitem}

\section{Regularization on the Particle Distribution}

In the non-regularized inversion experiments, the clustering is very likely to generate widely spreading particles at the depth where the data lose control. This can be improved by applying additional constraints that require concentrated distribution of the particles. I incorporate the measure of particle distribution into $\Phi_m$ in equation~\ref{eq:obj} so that the objective function is heavily penalized if the particles have wide distribution. The measure of distribution, referred to as the distribution norm or d-norm, can be any reasonable function that reflects the concentration of the particles. Given the locations of a group of particles, I consider two types of d-norm. The first is the variance, calculated as
\begin{equation}
\label{eq:variance}	
	\Phi_m = \frac{1}{n} \sum_{i=1}^{n} (z_i-\hat{z})^2,
\end{equation}
where $n$ is the number of particles, $z_i$ is the depth of the $i$th particle layer in 1D problem and $\hat{z}$ is the mean value of $z_i (i=1, \dots, n)$. The second is the range, calculated as
\begin{equation}
\label{eq:range}
	\Phi_m = max(z_i) - min(z_i), i  = 1, \dots, n.
\end{equation}
The following experiments will show the effect of using additional constraints of d-norm. The trade-off parameter $\gamma$ in equation~\ref{eq:obj} is set to a fixed value (0.01) so that $\Phi_d$ and $\Phi_m$ are numerically comparable.

\subsection*{Experiment 7: Perfect a priori information and using the variance constraint}

\begin{figure}
  \centering
    \includegraphics[width=0.7\textwidth]{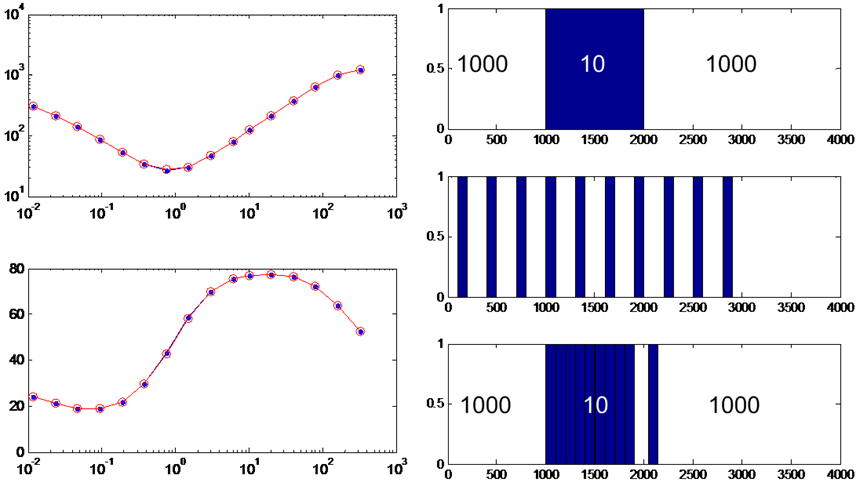}
  \caption{Experiment 7: perfect a priori information and using the variance constraint. }
  \label{fig:simple_variance}
\end{figure}

This experiment repeats the same scenario in Experiment 1, but the particles' locations are regularized by the d-norm function in equation~\ref{eq:variance} with the variance of the layer depths. Compared to Experiment 1, the target is recovered as a more compact layer while fitting the observed equally well (Figure~\ref{fig:simple_variance}). The deepest particle is detached from the main cluster, but the gap between them is not significant enough to cause much difficulty in the interpretation.

\subsection*{Experiment 8: Over-estimated resistivity and using the variance constraint}

\begin{figure}
  \centering
    \includegraphics[width=0.7\textwidth]{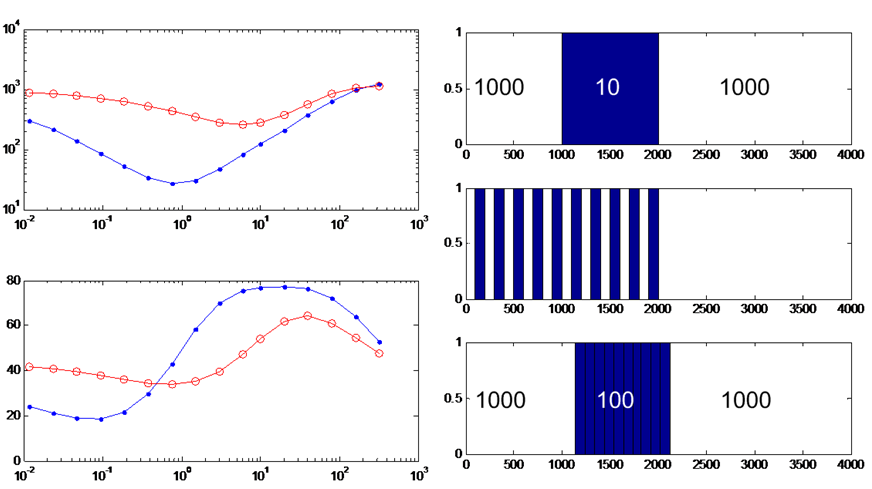}
  \caption{Experiment 8: over-estimated resistivity and using the variance constraint. }
  \label{fig:moreresistive_variance}
\end{figure}

I again test the imperfect a priori information with the use of variance d-norm constraint. To make the initial guess even worse, the target's resistivity (100 $\Omega$m) is assumed to be ten times greater than the true value (10 $\Omega$m). Unsurprisingly, the inversion cannot fit the data with inadequate conductive materials (Figure~\ref{fig:moreresistive_variance}). However, the recovered target is nicely compact, making the model still somewhat useful. In the struggle of fitting the data, the target is recovered at a slightly larger depth.

\subsection*{Experiment 9: Under-estimated resistivity and using the variance constraint}

\begin{figure}
  \centering
    \includegraphics[width=0.7\textwidth]{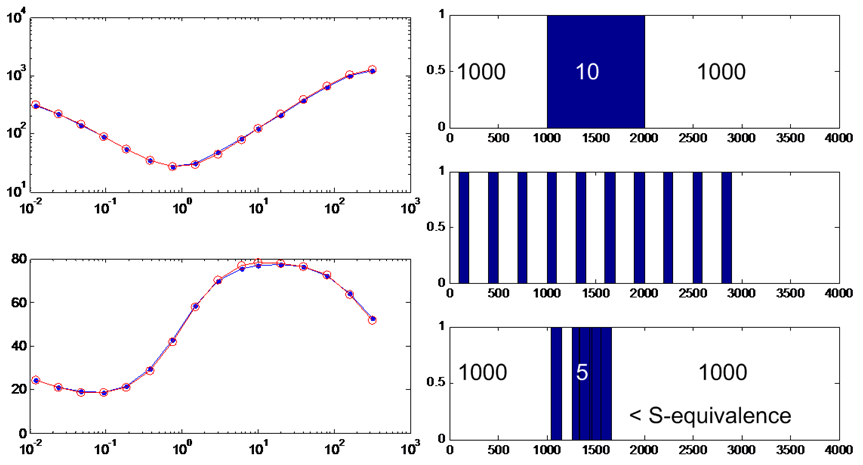}
  \caption{Experiment 9: under-estimated resistivity and using variance the constraint. }
  \label{fig:lessresistive_variance}
\end{figure}

Then I assume the target's resistivity is under-estimated to 5 $\Omega$m and examine the effect of the variance constraint on the layers' depths. With adequate conductive materials, the observed data can be fit perfectly. The excessive building materials in 5 particle layers are concealed by overlapping (Figure~\ref{fig:lessresistive_variance}). The layers' distribution concentrates to the depth between 1000 and 1700 m, a range narrower than the one recovered in Experiment 5.

\subsection*{Experiment 10: Multiple targets and using the variance constraint}

\begin{figure}
  \centering
    \includegraphics[width=0.7\textwidth]{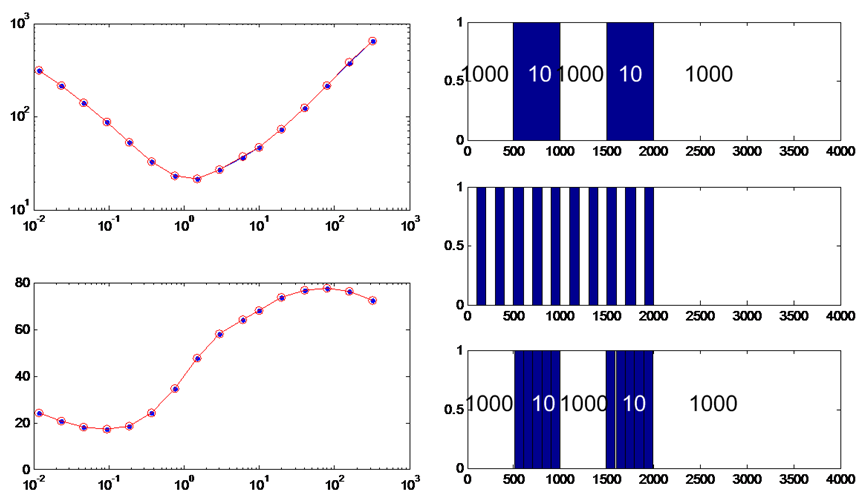}
  \caption{Experiment 10: multiple targets and using the variance constraint. }
  \label{fig:multilayers_variance}
\end{figure}

For multiple targets and perfect a priori information about the resistivity and total volume of the target, the additional variance constraint does a very good job confining the particle layers to a narrow band that coincides nicely with the true location of the target. The model recovered in Figure~\ref{fig:multilayers_variance} almost perfectly reconstructs the true model as the four boundaries of the two target layers are all delineated with high accuracy. 

\subsection*{Experiment 11: Perfect a priori information and using the range constraint}

\begin{figure}
  \centering
    \includegraphics[width=0.7\textwidth]{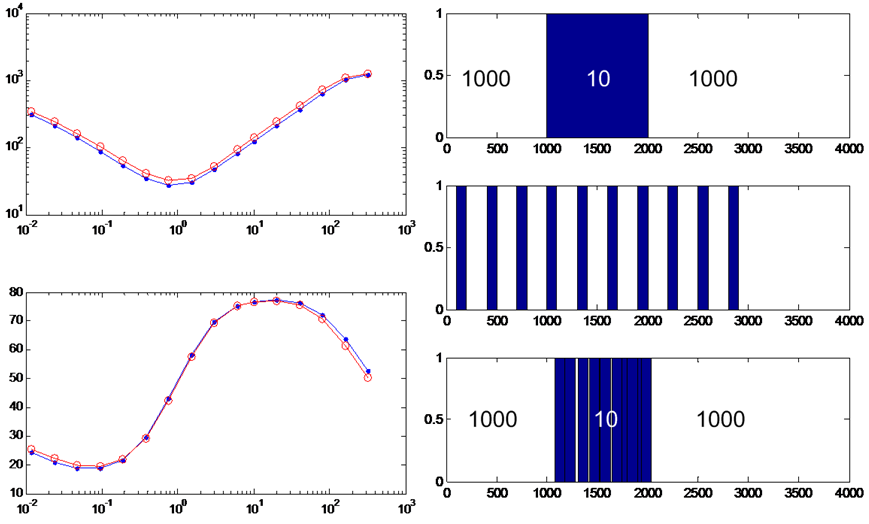}
  \caption{Experiment 11: perfect a priori information and using the range constraint. }
  \label{fig:simple_range}
\end{figure}

In this experiment, I test the perfect a priori information again but with the range constraint on the particle locations. As shown in Figure~\ref{fig:simple_range}, the particle layers cluster to the correct location of the target without significant detachment of any single layer. 

\subsection*{Experiment 12: Over-estimated resistivity and using the range constraint}

\begin{figure}
  \centering
    \includegraphics[width=0.7\textwidth]{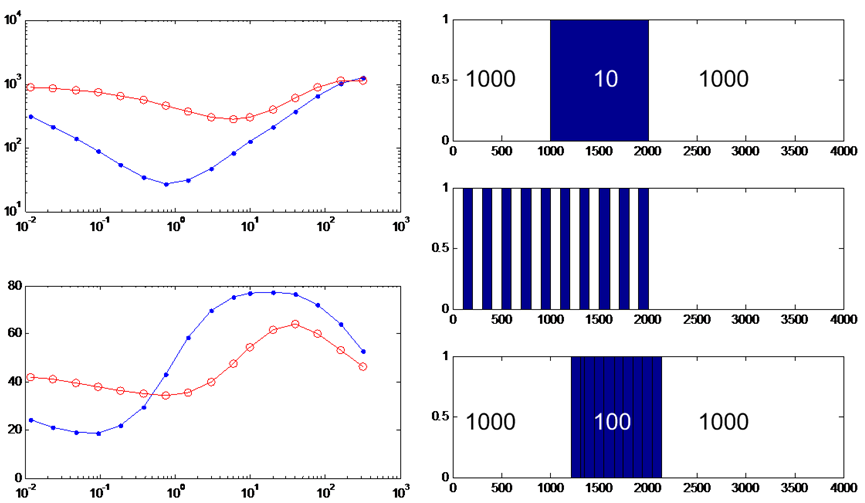}
  \caption{Experiment 12: over-estimated resistivity and using the range constraint. }
  \label{fig:moreresistive_range}
\end{figure}

While still using the range constraint, I test one of the imperfect scenarios where the target's resistivity is over-estimated to 100 $\Omega$m. Although the data misfit is really large, making it difficult to trust the inversion model, the recovered geometry of the target is not completely unreasonable (Figure~\ref{fig:moreresistive_range}). The similarity between Experiment 8 and Experiment 12 suggests the variance and range constraints have similar effect on bonding the particles together.

\subsection*{Experiment 13: Multiple targets and using the range constraint}

\begin{figure}
  \centering
    \includegraphics[width=0.7\textwidth]{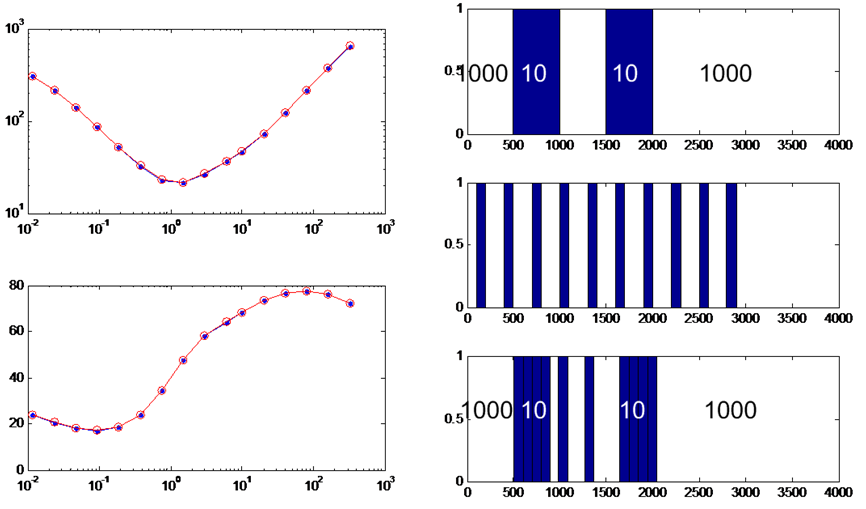}
  \caption{Experiment 13: multiple targets and using the range constraint. }
  \label{fig:multilayers_range}
\end{figure}

Finally I run a test with two targets and using the range constraint. The particle layers are clustered to the depth between 500 and 2000 m, a range corresponds well with the true model (Figure~\ref{fig:multilayers_range}). The recovered targets are not as accurate as in Experiment 10 with the variance constraint as the structure between the two targets is somewhat ambiguous. This can be explained by the insensitivity of the range norm in equation~\ref{eq:range} to the particle locations between the extremities. Nevertheless, the recovered model still indicates two separate target layers at different depths.

\section{Discussion}

\subsection*{Imperfect a priori information}
\begin{figure}
  \centering
    \includegraphics[width=0.5\textwidth]{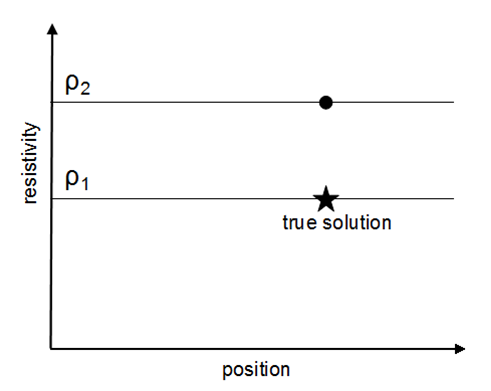}
  \caption{Particle clustering with imperfect a prioir information about the target. }
  \label{fig:optplot}
\end{figure}

As a tool to enforce strong a priori information about the known rock units, the particle clustering approach requires reasonably good initial guesses of the resistivity and the total volume of the target. In the 1D MT inversion examples presented above, there are two possible scenarios of imperfect information. The first is over-estimated total conductance, resulted from either under-estimated resistivity or too many particles, or both. This does not seem to cause serious problems since the unused particles can be effectively ignored if the particles are allowed to overlap (without superposition). The second scenario, in which not enough conductive materials are provided, has no chance to fit the data, but the recovered model may not be a bad representation of the true model. Figure~\ref{fig:optplot} explains why the inversion can be robust against the imperfect information. Assuming geometrically there is enough particles to resemble the target, a good recovery needs both of the resistivity and positions of the particles to be correct (at the star). If the information about the resistivity is correct ($\rho_1$), the true solution can be sought by adjusting the position; if the resistivity of the target is set to $\rho_2$, the model search can never reach the true solution, but the final result may be a point on the way approaching the true solution. Such a point, indicated by a dot in Figure~\ref{fig:optplot}, can be a model geometrically similar to the true solution.

Some adaptive algorithms can be developed to remedy the imperfect information. In the first scenario, unnecessary particles can be removed from the optimization resulting in reduced degree of freedom, if overlapping is detected. The second scenario can be possibly rescued by increasing the conductivity and/or adding more particles to the background. Allowing the a priori information to change would be the important step to make the algorithm practical.

\subsection*{Multiple clustering centers}

The two-layer examples have shown that the particle clustering is effective for more than one target, as the scattered particles successfully converged to the locations of the two target layers without any a priori information about the number of targets. However, the distribution constraint used in the multiple target example was identical to the one used in the single target example that encouraged all the particles to stay as close as possible. And computing the d-norm with the particles all together can be inappropriate and potentially problematic if many targets are far apart. This can be improved by either specifying the number of targets we want to recover beforehand or letting the inversion to decide whether the full set of particles should be divided into sub-groups if multiple clustering centers are appearing.  

\subsection*{Particle splitting in 2D/3D inverions}

The particle clustering behaved nicely for 1D layered earth because the location has only one variable, the depth. The inversion could become unstable in 2D/3D when the location is specified by two or three variables, in which case the relationship between the particles movement and the forward responses can be highly non-linear, and the inversion has the risk of being trapped in local minimum. The robustness of the particle clustering can be further compromised if too many particles are deployed leading to a large number of unknowns and severe ill-posedness. 

Here I propose a solution to this problem using adaptive particle size. Initially, there can be only one particle (e.g. a cubic prism) with its size as large as the target. The first few iterations find the optimal location of the single large particle that gives the best data fit. Once the data misfit cannot be reduced any more, the large particle is subdivided into four (in 2D) or eight (in 3D) equally-sized smaller particles, each of them are free to move independently from the original location of the large particle. The subdivision will repeat if the data misfit cannot be improved again. This process, using the same concept in the multi-level or multi-resolution approach, gradually increases the number of particles and expands the model space. The restriction on the model space at the initial stage makes the inversion more robust than starting with a large degree of freedom, and can be seen as another type of distribution constraint.

\section{Conclusions}

This paper presents the preliminary studies on a new approach of parameterizing a geophysical inversion model. My approach considers the model as some building blocks (particles) floating in a pre-determined background. The physical property of the particles is the a priori information one needs to assign before the inversion. Then the inversion searches the locations of the particles as the model parameters. The final model should have the particles clustered in a certain way that resembles the desired target. 

As the first pass, I tested the idea using 1D MT inversion as example. Numerical experiments, assuming both perfect and imperfect information about the target resistivity and number of particles, are carried out to demonstrate that the particle clustering approach is effective. I have found that the particle clustering is robust against incorrect a priori resistivity and number of particles even though the data are not acceptably fit, as the recovered models are still geometrically reasonable representation of the true models. A more complicated model containing two targets was also positively tested. 

Like other geophysical inversions, the incomplete information in the data necessitates the regularization on the particles' locations. Because usually a compact body is more geologically plausible, I propose a measurement, called distribution norm or d-norm, to encourage the particles to stick together while still fitting the data. By computing the variance or range of the locations, the d-norm penalizes the objective function heavily if the particles are widely spread. The use of the distribution constraints greatly tightened the particle clustering, yielding improved compactness to the recovered model. 

I also identify some practical problems in the particle clustering approach and the potential solutions, for example, how to correct the imperfect a priori information, clustering towards multiple centers to deal with multiple targets and stabilizing the optimization search in 2D/3D using adaptive particle splitting. Those are the necessary considerations that make the new approach practical and will be investigated in the future.


\bibliographystyle{seg}  
\bibliography{particle}

\end{document}